\documentclass[a4paper]{spie}  

\usepackage{amsmath,amsfonts,amssymb}
\usepackage{graphicx}
\usepackage[colorlinks=true, allcolors=blue]{hyperref}
\usepackage{color}
\usepackage{float}
\usepackage{tikz}
\usepackage{gensymb}
\usepackage{printlen}
\usepackage{changes} 
\usetikzlibrary{shapes.geometric, arrows}
\tikzstyle{startstop} = [rectangle, rounded corners, minimum width=3cm, minimum height=1cm,text centered, draw=black, fill=red!30]
\tikzstyle{io} = [trapezium, trapezium left angle=80, trapezium right angle=100, minimum width=2cm, minimum height=1cm, text centered, draw=black, fill=blue!30]
\tikzstyle{process} = [rectangle, minimum width=3cm, minimum height=1cm, text centered, draw=black, fill=orange!30]
\tikzstyle{decision} = [rectangle, minimum width=3cm, minimum height=1cm, text centered, draw=black, fill=green!30]
\tikzstyle{arrow} = [thick,->,>=stealth]

\title{The on-ground data reduction and calibration pipeline for SO/PHI-HRT}

\author[1]{J. Sinjan}
\author[1]{D. Calchetti}
\author[1]{J.~Hirzberger}
\author[2]{D.~Orozco Su\' arez}
\author[1]{K.~Albert}
\author[1]{N. Albelo Jorge}
\author[3]{T.~Appourchaux} 
\author[4]{A.~Alvarez-Herrero}
\author[5]{J.~Blanco Rodr\'\i guez}
\author[1]{A.~Gandorfer}
\author[1]{D.~Germerott}
\author[1]{L.~Guerrero}
\author[1]{P.~Gutierrez Marquez}
\author[1]{F. Kahil}
\author[1]{M.~Kolleck}
\author[1,10]{S.K.~Solanki}
\author[2]{J.C.~del Toro Iniesta}
\author[6]{R.~Volkmer}
\author[1]{J.~Woch}
\author[7]{B.~Fiethe}
\author[8]{J.M.~Gómez Cama}
\author[9]{I.~P\' erez-Grande}
\author[5]{E.~Sanchis Kilders}
\author[2]{M.~Balaguer Jiménez}
\author[2]{L.R.~Bellot Rubio}
\author[8]{M.~Carmona}
\author[1]{W.~Deutsch}
\author[1,9]{G.~Fernandez-Rico}
\author[4]{A.~Fern\' andez-Medina}
\author[4]{P.~Garc\'\i a Parejo}
\author[5]{J.L.~Gasent Blesa}
\author[1,11]{L.~Gizon}
\author[1]{B.~Grauf}
\author[1]{K.~Heerlein}
\author[1]{A.~Lagg}
\author[7]{T.~Lange}
\author[2]{A.~L\' opez Jim\' enez}
\author[6]{T.~Maue}
\author[1]{R.~Meller}
\author[7]{H.~Michalik}
\author[2]{A.~Moreno Vacas}
\author[1]{R.~M\" uller}
\author[6]{E.~Nakai}
\author[6]{W.~Schmidt}
\author[1]{J.~Schou}
\author[1]{J.~Sch\"uhle}
\author[1]{J.~Staub}
\author[2]{H.~Strecker}
\author[9]{I.~Torralbo}
\author[1]{G.~Valori} 

\affil[1]{Max-Planck-Institut für Sonnensystemforschung, Justus-von-Liebig-Weg 3, 37077 Göttingen, Germany}
\affil[2]{Instituto  de  Astrofísica  de  Andalucía  (IAA-CSIC),  Apartado  deCorreos 3004, E-18080 Granada, Spain}
\affil[3]{Univ.  Paris-Sud,  Institut  d’Astrophysique  Spatiale,  UMR  8617,CNRS, Bâtiment 121, 91405 Orsay Cedex, France}
\affil[4]{Instituto Nacional de Técnica Aeroespacial, Carretera de Ajalvir, km 4, E-28850 Torrejón de Ardoz, Spain}
\affil[5]{Universitat   de   València,   Catedrático   José   Beltrán   2,   E-46980 Paterna-Valencia, Spain}
\affil[6]{Leibniz-Institut   für   Sonnenphysik,    Schöneckstr.    6,    D-79104 Freiburg, Germany}
\affil[7]{Institut für Datentechnik und Kommunikationsnetze der TU Braunschweig, Hans-Sommer-Str. 66, 38106 Braunschweig, Germany}
\affil[8]{University of Barcelona, Department of Electronics, Carrer de Martíi Franquès, 1 - 11, 08028 Barcelona, Spain}
\affil[9]{Instituto  Universitario  "Ignacio  da  Riva",  Universidad  Politécnicade Madrid, IDR/UPM, Plaza Cardenal Cisneros 3, E-28040 Madrid, Spain}
\affil[10]{School   of   Space   Research,   Kyung   Hee   University,   Yongin, Gyeonggi-Do, 446-701, Korea}
\affil[11]{Institut   für   Astrophysik,   Georg-August-Universität   Göttingen, Friedrich-Hund-Platz 1, 37077 Göttingen, Germany}

\authorinfo{Further author information: (Send correspondence to J.Sinjan)\\J.Sinjan: E-mail: sinjan@mps.mpg.de}

\begin{document} 
\maketitle

\begin{abstract}
The ESA/NASA Solar Orbiter space mission has been successfully launched in February 2020. Onboard is the Polarimetric and Helioseismic Imager (SO/PHI), which has two telescopes, a High Resolution Telescope (HRT) and the Full Disc Telescope (FDT). The instrument is designed to infer the photospheric magnetic field and line-of-sight velocity through differential imaging of the polarised light emitted by the Sun. It calculates the full Stokes vector at $6$ wavelength positions at the Fe I $617.3$ nm absorption line. Due to telemetry constraints, the instrument nominally processes these Stokes profiles onboard, however when telemetry is available, the raw images are downlinked and reduced on ground. Here the architecture of the on-ground pipeline for HRT is presented, which also offers additional corrections not currently available on board the instrument. The pipeline can reduce raw images to the full Stokes vector with a polarimetric sensitivity of $10^{-3}\cdot I_{c}$ or better. 
\end{abstract}


\section{Introduction}
\label{sec:intro}
The Solar Orbiter (SO) is the first selected medium-class mission from ESA's Cosmic Vision 2015-2025 Program. It is a collaborative effort with NASA, and the spacecraft was launched on February 10th 2020\cite{Muller2020TheOverview}. Its primary goal is to study the Sun and the inner heliosphere. To achieve this, the spacecraft carries a scientific payload of 10 instruments, 6 remote sensing and 4 in-situ. The Polarimetric and Helioseismic Imager (PHI) is one of the remote sensing instruments and it retrieves the continuum intensity, the vector magnetic field, and the line-of-sight velocity, both in the Sun's photosphere\cite{solanki_polarimetric_2020}. It does this by sampling four linear combinations of the four Stokes parameters of the magnetically sensitive Fe I absorption line at $617.3$ nm at 6 wavelength positions. These samples are later transformed to the aforementioned physical quantities by inverting the radiative transfer equation for polarized light in the presence of a magnetic field.

The instrument has two telescopes: the Full Disc Telescope (FDT) and the High Resolution Telescope (HRT)\cite{gandorfer_high_2018}. The FDT can capture the full solar disc, while the HRT is designed to capture photospheric features in more spatial detail. During normal operations to conserve telemetry, the raw images from the two telescopes are reduced onboard using field-programmable gate array (FPGA) computers, producing the data products which are downlinked to Earth\cite{Lange2017On-boardInstrument, Albert2020}.

However there are occasions during the nominal and extended mission phases when telemetry rates will be favourable such that raw images can be downloaded and reduced on-ground. The ability to reduce data on-ground provides more opportunities for fine-tuning the data reduction process, and therefore for producing higher quality data products. Furthermore it allows for the additional processing when the HRT's stabilisation system is switched off. Since launch the spacecraft has been in its cruise phase until 27th November 2021, when the nominal mission phase began\cite{Zouganelis2020}. During the cruise phase, predominantly raw images from HRT were downloaded and reduced on-ground, to allow for in-flight testing and investigate the instrument's performance. This paper outlines the current state of the on-ground pipeline for the HRT (V1.4 June 2022), written in Python3, which is used to reduce and calibrate these raw images. Finally early in-flight data reduced with this pipeline is presented and the telescope's high polarimetric sensitivity is shown. 


\section{SO/PHI-HRT}

\subsection{The Instrument}
\label{subsec:hrt_inst}
\begin{figure} [H]
\includegraphics[trim=0.2cm 4cm 0cm 0.2cm,clip,width = \textwidth]{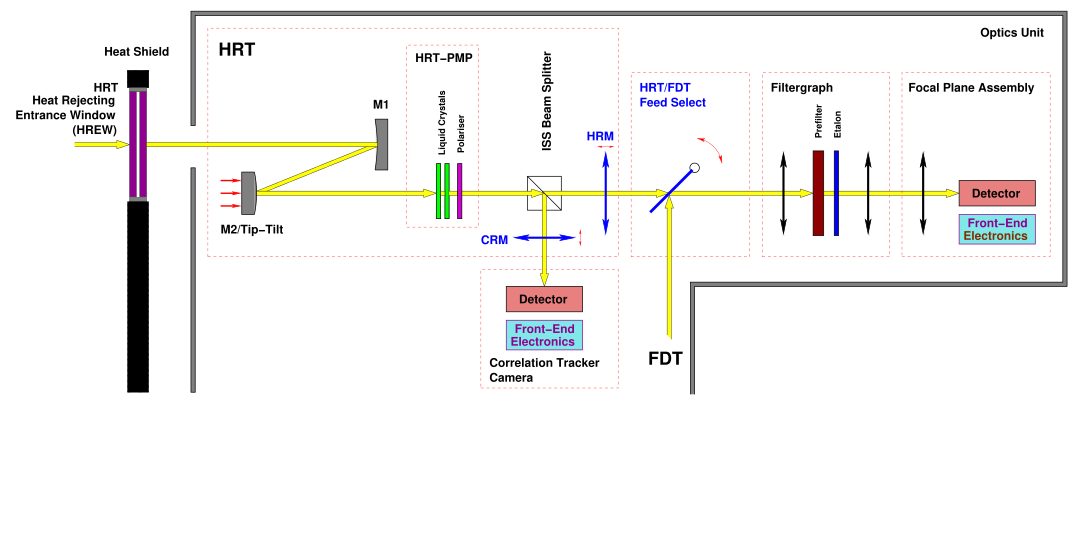}
\caption[schematic] 
{ \label{fig:hrt} Schematic of the optics unit of SO/PHI-HRT. SO/PHI-FDT is located in the same optics unit but is not shown for clarity.}
\end{figure}

A simplified description of the key optical components of the HRT is presented, following the optical path shown in Fig. \ref{fig:hrt}\cite{solanki_polarimetric_2020, gandorfer_high_2018}. The HRT is a Ritchey-Chrétien telescope with a decentred pupil. The HRT has a Heat Rejection Entrance Window (HREW) that allows in only $4\%$ of the incoming solar light power. The HRT uses its own Polarisation Modulation Package (PMP), consisting of Liquid Crystal Variable Retarders (LCVR) and a polariser to modulate the light in order to obtain the polarisation characteristics of the incoming light\cite{Alvarez-Herrero2015a}. Once modulated, the light is split and a portion enters the Correlation Tracker Camera (CTC).  The CTC is part of the Image Stabilisation System (ISS) which works to track a specific feature on the solar surface, calculating from the CTC images the steering signal for the M2 (tip-tilt) mirror, to accurately track the desired feature\cite{Volkmer2012, Carmona2014}. The ISS is also used to compensate for effects such as spacecraft jitter. After the beam splitter the light goes through the HRT Refocus Mechanism (HRM) and passes through the Feed Select Mechanism (FSM), which is used to switch between FDT and HRT, and towards the Filtergraph (FG). The FG contains two pre-filters and a tunable LiNbO$_{3}$ Fabry-Perot etalon, an interferometer, which together allow for a transmission window with a mean full-width-half-maximum (FWHM) of $(106\pm5)$ m\AA $\:$and free spectral range of $3.0$ \AA \cite{gandorfer_high_2018, Dominguez-Tagle2014OpticalPHI}. The resultant light then illuminates an Active Pixel Sensor (APS) which reads out images with $2048\times2048$ pixels. The re-imaging optics in the Focal Plane Assembly (FPA) provides a plate scale on the sensor of $0.5^{\prime\prime}$ per pixel which corresponds to $102$ km at $0.28$\,AU distance.

The imaging cadence is controlled by the modulation accumulation scheme in the PMP. While the HRT has the operational capability of a $60$ second cadence, during the cruise phase a $96$ second cadence was used. This was done using a PMP scheme of $[4,5]$. At each wavelength position, $4$ frames are taken for each modulation state, and this is cycled through each modulation state $5$ times, resulting in $20$ total frames for each modulation state at each wavelength position. The minimum number of frames to achieve the desired signal to noise ratio of $10^{-3}$ is $16$\cite{solanki_polarimetric_2020}.

\subsection{Flat-Field Acquisition}\label{subsec:flat_acq}
A critical reduction process of the science data requires a flat-field. The flat-field contains the difference in gain of a given pixel with respect to the others, as well as information of other imperfections such as dust grains in the field of view (FOV). The HRT flat-fields are not acquired using off-pointing of the spacecraft\cite{kuhn_gain_1991}. Instead, the solar surface evolution, is used to introduce differences between subsequent images, such that localised solar features are averaged out with enough acquisitions. Over a period of approximately $8$ hours, 1500 images are accumulated at each polarisation state and wavelength. These flat-fields are acquired during every major campaign, to ensure the science data can be properly calibrated. However polarimetric structures remain in the flat-field that are smeared horizontally due to the solar rotation. This horizontal smearing leaves unwanted artefacts when applying the flat-field correction to the scientific data. Therefore an additional flat-field processing procedure was implemented as part of the pipeline: unsharp masking, which is described in Sec. \ref{subsec:unsharp_mask}. 


\section{SO/PHI-HRT On ground Pipeline}
\subsection{Pipeline Overview}\label{subsec:pipe}
\begin{figure} [H]
   \begin{center}
   \begin{tabular}{c} 
   \includegraphics[trim={2cm 0cm 2cm 4cm},clip,height=8.6in]{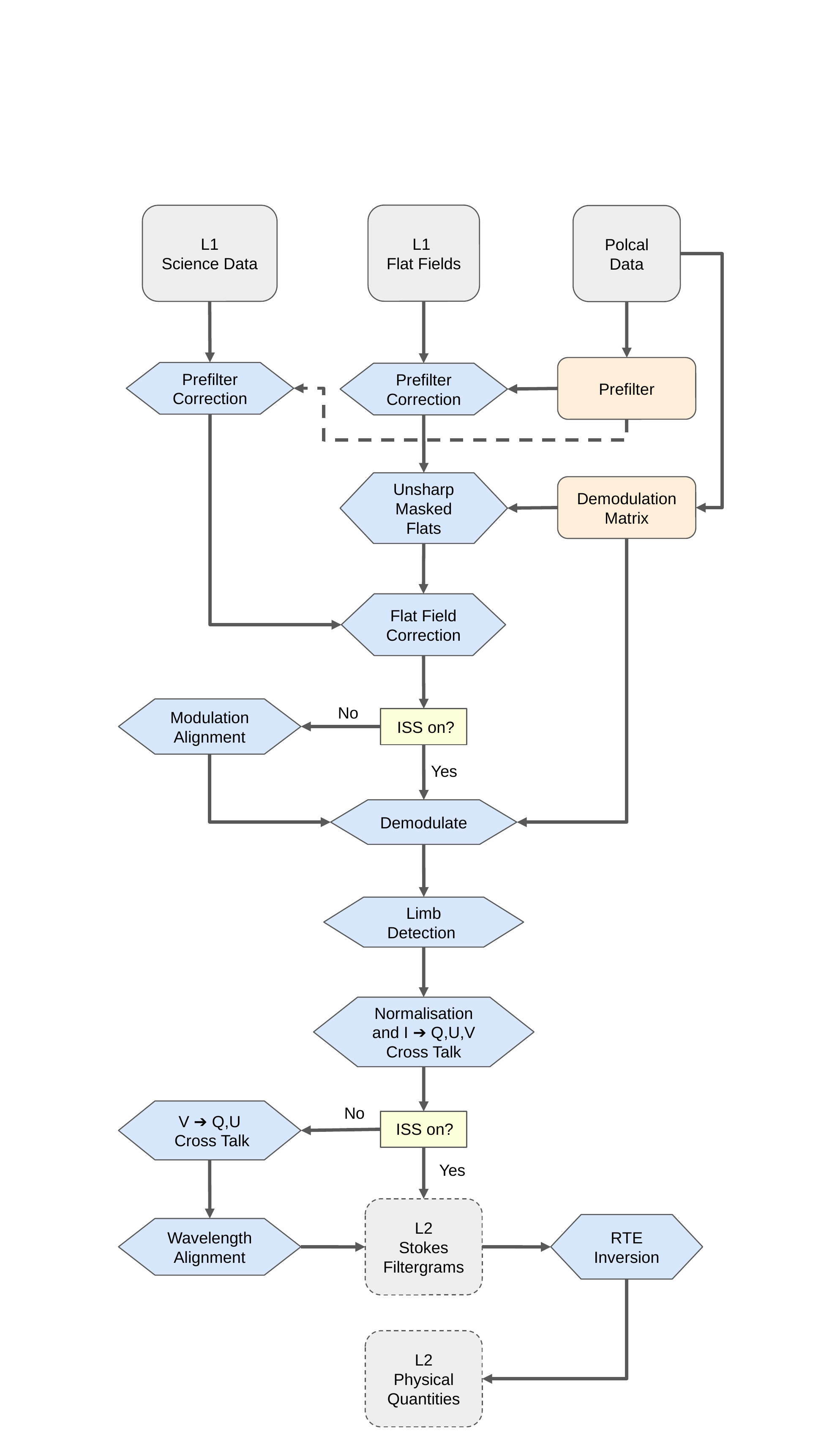}
   \end{tabular}
   \end{center}
   \caption[flowchart] 
    { \label{diag:pipe} Flowchart of the on-ground pipeline, reducing Level 1 data to the Level 2 physical quantities. The dark field correction and field stop application has been omitted for brevity. `Polcal' refers to the on-ground polarisation calibration campaign prior to launch.}
   \end{figure}

The on-ground pipeline is developed in Python3 and reduces the raw data received from the SO/PHI-HRT instrument. The raw files downloaded from the instrument are classed Level 0 (L0) data. They become Level 1 (L1) data once necessary metadata are added, the data are scaled to the correct units (to account for the compression scheme used) and reflected in the $Y$ axis to match the solar orientation convention. This on-ground pipeline converts the L1 data, into L2 data. This process is described in Fig. \ref{diag:pipe}.

The inputs to the pipeline are the science data, raw flat-fields, and demodulation matrices (for each operating PMP temperature) from the polarisation calibration campaign performed prior to launch. The science data and flat-fields are first dark-corrected to remove the dark current (not shown in Fig. \ref{diag:pipe} for brevity). A key capability of the pipeline is the option to unsharp mask the flat-fields (see Sec. \ref{subsec:unsharp_mask}), however the width of the Gaussian distribution to be used must be known beforehand. The pipeline has the functionality to reduce multiple datasets at once, with the same flat-field, provided the image dimensions, PMP temperature and continuum position of all datasets agree. The pipeline is built to work with any cropped dataset provided the input is square. The pipeline can also reduce images when the ISS is locked to track the solar limb, using an automatic limb detection algorithm to account for limb darkening effects when normalising the Stokes parameter and applying the cross-talk correction. Furthermore additional steps are implemented to account for the case when the ISS is not operating. The output quantities are indicated with the dotted outlines in Fig. \ref{diag:pipe}.

\subsection{Unsharp Masking}\label{subsec:unsharp_mask}
\begin{figure} [H]
\begin{center}
\includegraphics[width = \textwidth]{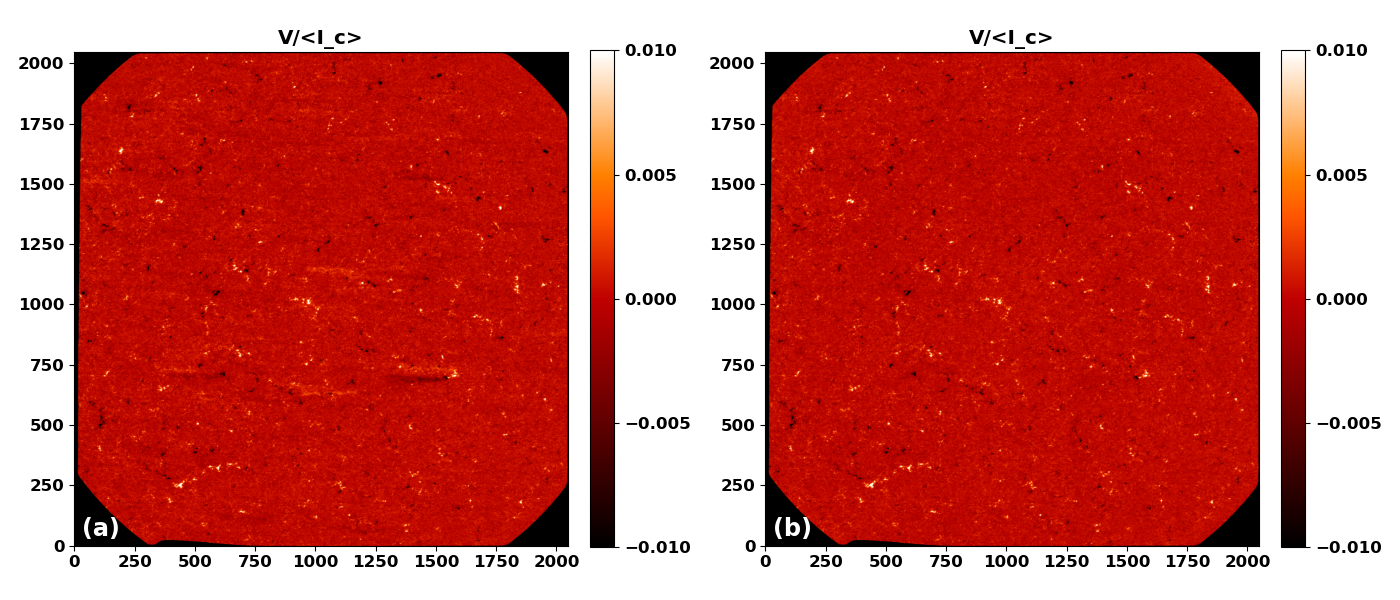}
\caption[unsharpmask] 
{ \label{fig:unsharp} A Stokes $V$ image in the line wing on 23 February 2021 17:00 UTC, taken at a Sun-spacecraft distance of $0.526$ AU: a) result when using a flat-field without unsharp masking, b) result when unsharp masking is applied. The region outside the field stop is set to black for clarity.}
\end{center}
\end{figure}

Due to the method of flat-field acquisition as described in Sec. \ref{subsec:flat_acq}, horizontal polarisation elements exist in the flat-fields, which would contaminate the data. To remove this contamination unsharp masking is performed on the flat-fields. This is achieved by convolving the demodulated flat-fields with a 2D Gaussian distribution. The width of the Gaussian distribution was optimised such that the horizontal stripes were removed, but that any larger scale information was retained. The width of the Gaussian is a function of Solar distance and PMP temperature. For example, at a distance of $0.526$ AU, a width of $59$ pixels was used, while at $0.801$ AU, $49$ pixels was the appropriate width to be used. An example of the unsharp masking process is shown in Fig. \ref{fig:unsharp}.

\subsection{Flat-Field Correction}\label{subsec:flat_corr} 
From in-flight testing it was determined that a polarisation state dependent, and wavelength dependent flat-field must be applied to the science data. This is done in order to remove a polarimetric ghost that was detected, likely to originate from a reflection between the inner panel of the HREW and the highly reflective etalon. With an optical path of this nature, with many parallel optical surfaces, several measures were taken to suppress ghosts; however it was not possible to eliminate all of them. In particular those produced by the HREW, which is not mounted on the instrument but instead is a component of the heat shield of the spacecraft is prone to have large margins in the mechanical alignment. Nevertheless, with the proper treatment of the flat-fields, we are capable of removing the contribution from the detected ghost to below the noise requirement level. Finally, to correct for the cavities within the etalon, the flat-field must be normalised over the wavelength range, so that the spectral line profile is removed. This also has the effect of removing the solar rotation to at least first order. Thus the flat-fielded data, $I_{ff}$, is calculated as follows:
\begin{equation}
    I_{ff}(x,y,s,\lambda) = \frac{I_{df}(x,y,s,\lambda)}{I_{flat}(x,y,s,\lambda)},
\end{equation}
where $I_{df}$ is the dark-corrected data, $x$ and $y$ are the spatial dimensions in the FOV, $s$ is the polarization state and $\lambda$ denotes the wavelength.

\subsection{Demodulation}\label{subsec:demod}
\begin{figure} [H]
\begin{center}
\includegraphics[width = \textwidth]{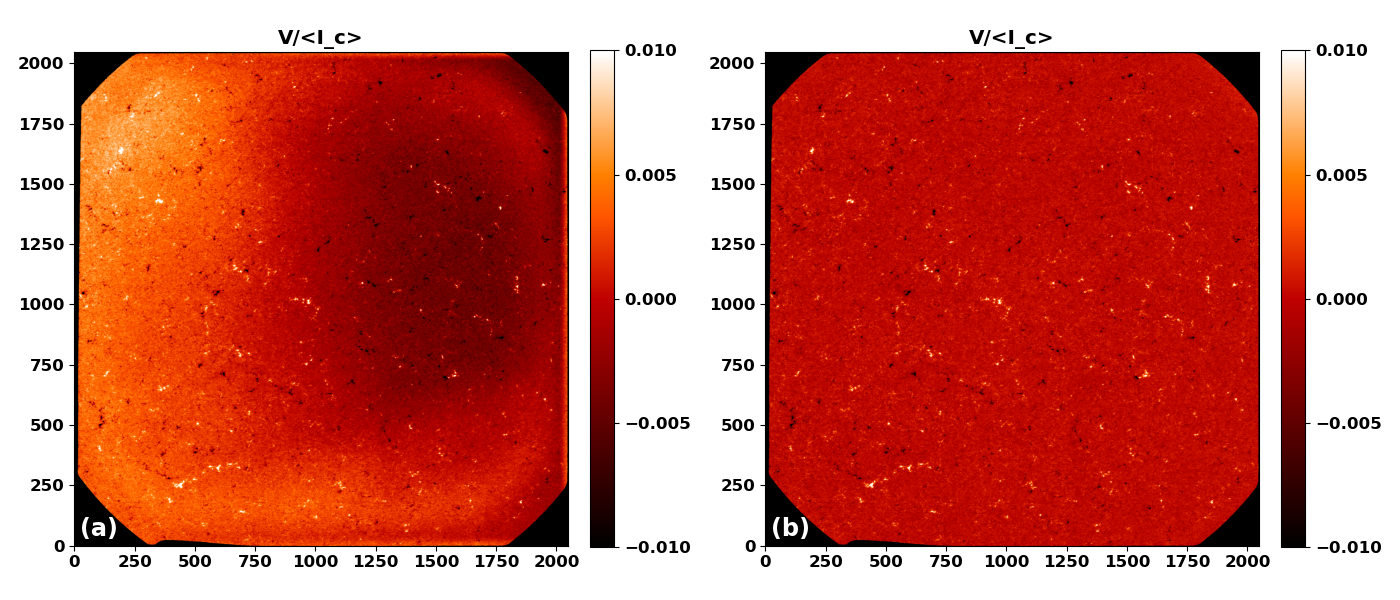}
\caption[demod] 
{ \label{fig:demod} A Stokes $V$ image in the ling wing on 23rd February 2021 17:00 UTC, with a distance of $0.526$ AU and PMP temperature of $50$ $\degree$C: a) result when using the original demodulation matrix; b) result after using an averaged matrix.}
\end{center}
\end{figure}

The raw images must be demodulated to remove the modulation applied by the PMP in the image acquisition process. This is obtained with the demodulation matrix: $d_{11-44}$, for a given pixel:

\begin{equation}\label{mod_demod}
\begin{pmatrix}
d_{11} & \dots & d_{14}\\
\vdots & \ddots & \\
d_{41} & & d_{44}
\end{pmatrix}
\begin{pmatrix}
I_{ff1}(x,y,s,\lambda)\\
I_{ff2}(x,y,s,\lambda)\\
I_{ff3}(x,y,s,\lambda)\\
I_{ff4}(x,y,s,\lambda)
\end{pmatrix} = 
\begin{pmatrix}
S_{1}\\
S_{2}\\
S_{3}\\
S_{4}
\end{pmatrix},
\end{equation}
where $I_{1...4ff}$ are the flat-fielded intensities and $S_{1...4}$ are the Stokes parameters: $I,Q,U,V$.
The demodulation matrices acquired during the on-ground testing before launch, resulted in large gradients across the field of view in the Stokes parameters, an example of which are portrayed in Fig. \ref{fig:demod}. Averaging the central $1024\times1024$ region of the demodulation matrix results in a demodulation matrix that removes the large scale FOV variations that were introduced using the matrix measured during the on-ground polarisation campaign.

\subsection{Limb Detection and Normalisation}\label{subsec:limb_norm} 
For datasets where the limb is in the FOV, such as when the spacecraft is off-pointing to the poles, certain additional steps are needed. From the World Coordinate System (WCS) information in the fits header, the pointing (North, South, East, West) is determined, such that a limb fitting algorithm can accurately detect the limb. First a mask is created, ensuring that all pixels outside the solar disc are set to $0$ in the final data products. To prevent limb darkening from affecting the normalisation, the edge and radius of the limb are calculated. For limb images, the average of Stokes $I$ at the continuum wavelength position is used as the Stokes normalisation factor, making sure to only include pixels which are less than $80\%$ of the solar radius in distance from disc centre. Under disc centre pointing, the average from the central $1024\times1024$ region is found and used as $I_{c}$.

\subsection{Cross-Talk Correction}\label{subsec:ctalk} 
Cross-talk between the Stokes parameters arises from three main sources: spacecraft jitter, imperfect instrument calibration, and modulation from the LCVRs. The strongest cross-talk, is that from Stokes $I$ to the other Stokes parameters, as the absolute value of Stokes $I$ is much greater than that of $Q,\;U,\;V$\cite{DelToroIniesta2003IntroductionSpectropolarimetry}. Due to cross-talk from sources described earlier, an ad-hoc correction is applied to the data\cite{SanchezAlmeida1992ObservationSunspots, schlichenmaier_spectropolarimetry_2002}. A linear fit of $Q,\;U,\;V$ against $I$ is performed separately, on the continuum wavelength image, to find the gradient and offset parameters of the cross-talk from $I$ to $Q,\;U,\;V$. When applying the cross-talk correction  at each of the $6$ different wavelength positions, the parameters are weighted by the respective averaged Stokes $I$ value, relative to the continuum value. The cross-talk parameters from in-flight data are of the order of $1\%$ or lower, indicating that the ISS ensures there are no major contributions from the spacecraft jitter, the instrument calibration is accurate and that the demodulation matrices used are effective. After this step, provided the ISS is operational, the pipeline produces the L2 `Stokes' filtergrams.

\subsection{Special Case: ISS Off}\label{subsec:iss}
The ISS of the instrument, as explained in Sec. \ref{subsec:hrt_inst}, tracks features on the Sun and compensates for the spacecraft jitter. The latter is important for two reasons: it removes the cross-talks induced by the jitter and keep the 24 raw frames aligned between each other during the acquisition. In some occasions the ISS has to be turned off and three procedures have been implemented to compensate for the absence of this subsystem.
\subsubsection{Modulation Alignment}\label{subsubsec:mod}
\begin{figure} [H]
\begin{center}
\includegraphics[width = \textwidth]{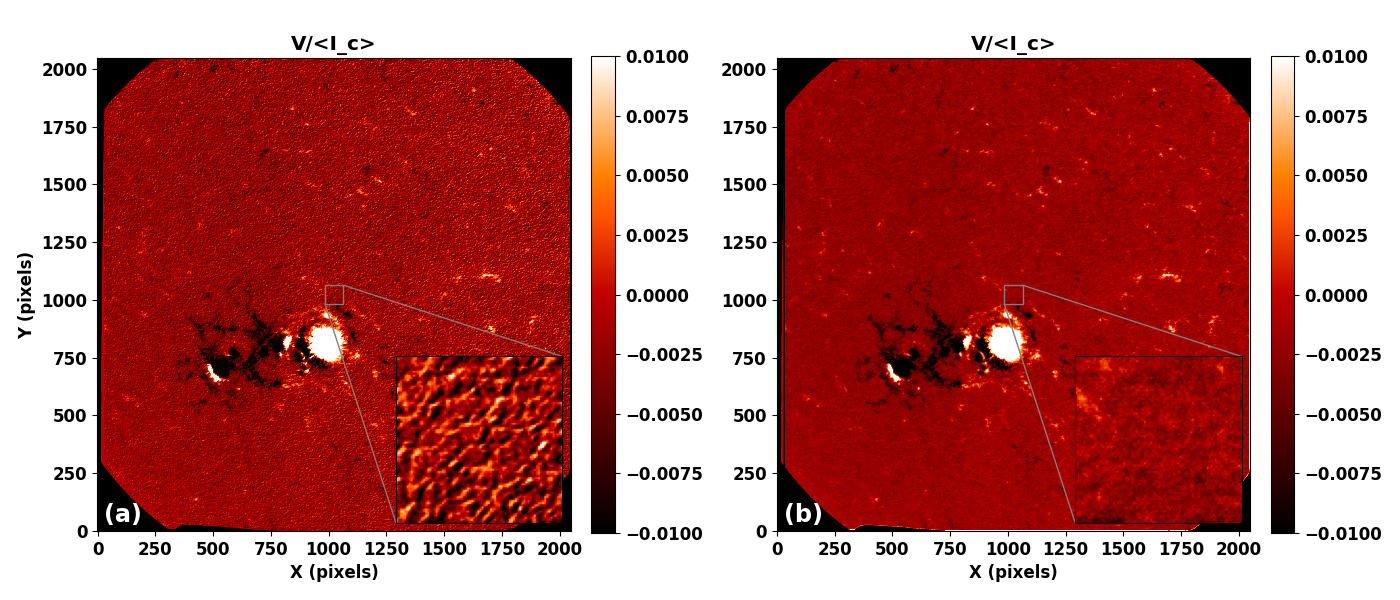}
\caption[mod align] 
{ \label{fig:mod_align} A Stokes $V$ image in the line wing on 7 March 2022 00:00 UTC, at a Sun-spacecraft distance of $0.501$ AU and PMP temperature of $50 \degree$C: a) result when using the standard pipeline procedure, b) result when using the modulation alignment step.}
\end{center}
\end{figure}

The first procedure is the modulation alignment just before the demodulation of the data (see Fig. \ref{diag:pipe}). For each wavelength we consider the first polarimetric modulation as a reference; the remaining three polarimetric modulations are then aligned to the chosen reference. This is performed by computing the gradient of the images, selecting a sub-region of $512\times512$ pixels, and evaluating the cross correlation between them with sub-pixel accuracy\cite{Guizar-Sicairos2008}. This registration has to be performed before the demodulation in order to avoid the combination of pixels from different regions on the Sun. Figure \ref{fig:mod_align} shows the effect of the spacecraft jitter on the data and the removal of the noise pattern with the modulation alignment step. The Stokes $V$ noise level decreases from $2.4\times10^{-3}$ to $1.4\times10^{-3}$.
\subsubsection{\textit{V} to \textit{Q,U} Cross-Talk Correction}\label{subsubsec:ctalk}
The spacecraft jitter is responsible for increasing the cross-talk both from $I$ to $Q,\;U,\;V$ and from $V$ to $Q,\;U$. Similar to the correction of cross-talk from Stokes $I$ (Sec. \ref{subsec:ctalk}), this procedure performs a linear fit of Stokes $Q$ and $U$ against Stokes $V$ immediately after the cross-talk correction from Sec. \ref{subsec:ctalk} is applied. The difference between the two methods is that here we consider points from all the wavelengths while computing the linear fit, so the parameters are not weighted by the continuum value. Despite the cross-talk parameters from $I$ to $Q,\;U,\;V$ being of the order of $1\%$, the parameters from $V$ to $Q,\;U$ can be up to $7\%$. 

\subsubsection{Wavelength Alignment}\label{subsubsec:wl}
The last step before producing the L2 `Stokes' filtergrams is the alignment of the frames at different wavelengths. Similar to that described in Sec. \ref{subsubsec:mod}, we use the continuum Stokes $I$ image as a reference, and after computing the gradient, we align the other frames to this reference. The only exception is for the line core wavelength image, for which a line wing image is used as a reference. This alignment has to be performed before the Radiative Transfer Equation is inverted to create cohesive Stokes profiles, where the different wavelength samples in a particular Stokes profile, come from the same spatial location on the Sun. 
\subsection{Radiative Transfer Equation Inversion}\label{subsec:rte}
 To infer the physical quantities from the Stokes maps, a Radiative Transfer Equation (RTE) inversion is performed. Similar to the inversion code used by the HMI vector magnetic field pipeline\cite{hoeksema_helioseismic_2014, Borrero2011}, a code assuming a Milne-Eddington (ME) atmosphere is used\cite{DelToroIniesta2003IntroductionSpectropolarimetry,LandiDeglInnocenti2005}. A ME model assumes that the physical properties of the atmosphere remain constant with geometrical height, while the source function scales linearly with optical depth. 
 
 This pipeline uses the CMILOS code written in C, which utilises analytical response functions \cite{orozco_suarez_usefulness_2007}. This code is the same as that used by the FPGA devices onboard\cite{CobosCarrascosa2016} and it works by minimizing the difference between the observed and synthetic profiles it produces, iterating the atmosphere's parameters until convergence of the two profiles is achieved. The CMILOS code has three operating modes:
 \begin{itemize}
     \item RTE with default starting conditions
     \item RTE with Classical Estimates as starting conditions
     \item Classical Estimates only
 \end{itemize}
 With Classical Estimates (CE) enabled, either in CE only mode, or together with RTE, it estimates the line-of-sight magnetic field and velocity using the centre of gravity method\cite{Semel1967,Rees1978}. The transverse component of the magnetic field is estimated using the weak-field approximation\cite{LandiDeglInnocenti2005}. The CMILOS inversion code produces the following L2 data products: full magnetic vector, Dopplergram and continuum intensity. The azimuth is defined as the counter-clockwise rotation from the positive direction of the detector $y$-axis. However, the intrinsic $180^\circ$ ambiguity of the Zeeman effect is not removed at this stage.

\section{Early In-Flight Data} 
\label{sec:data}
\subsection{February 2021}
We first introduce reduced data from 23 February 2021 17:00 UTC, during the short term planning period (STP) 136. This data captured the quiet Sun at disk centre, allowing us to characterise the noise level well, given the lack of strong magnetic field signals. The distance of the spacecraft to the Sun was $0.526$ AU, and the PMP temperature was set to $50\;\degree$C.  At this distance, the (two-pixel) spatial resolution is $382$ km.

\begin{figure} [H]
\begin{center}
\includegraphics[width = \textwidth]{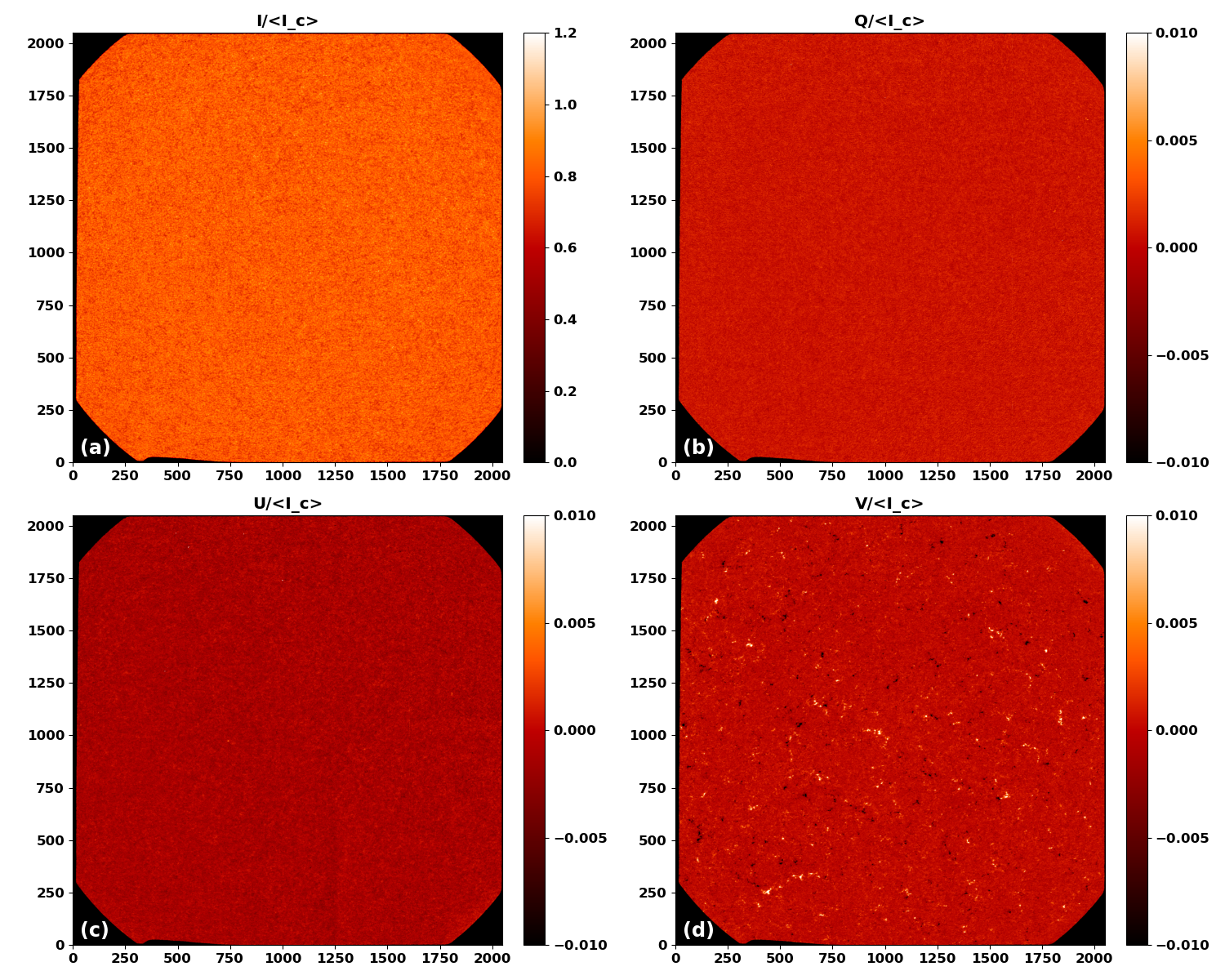}
\caption[stokes] 
{ \label{fig:iquv}Stokes filtergram in the red line wing on 23 February 2021 17:00 UTC: a) Stokes $I/I_{c}$, b) Stokes $Q/I_{c}$, c) Stokes $U/I_{c}$, d) Stokes $V/I_{c}$.}
\end{center}
\end{figure}

\begin{figure} [H]
\begin{center}
\includegraphics[width = \textwidth]{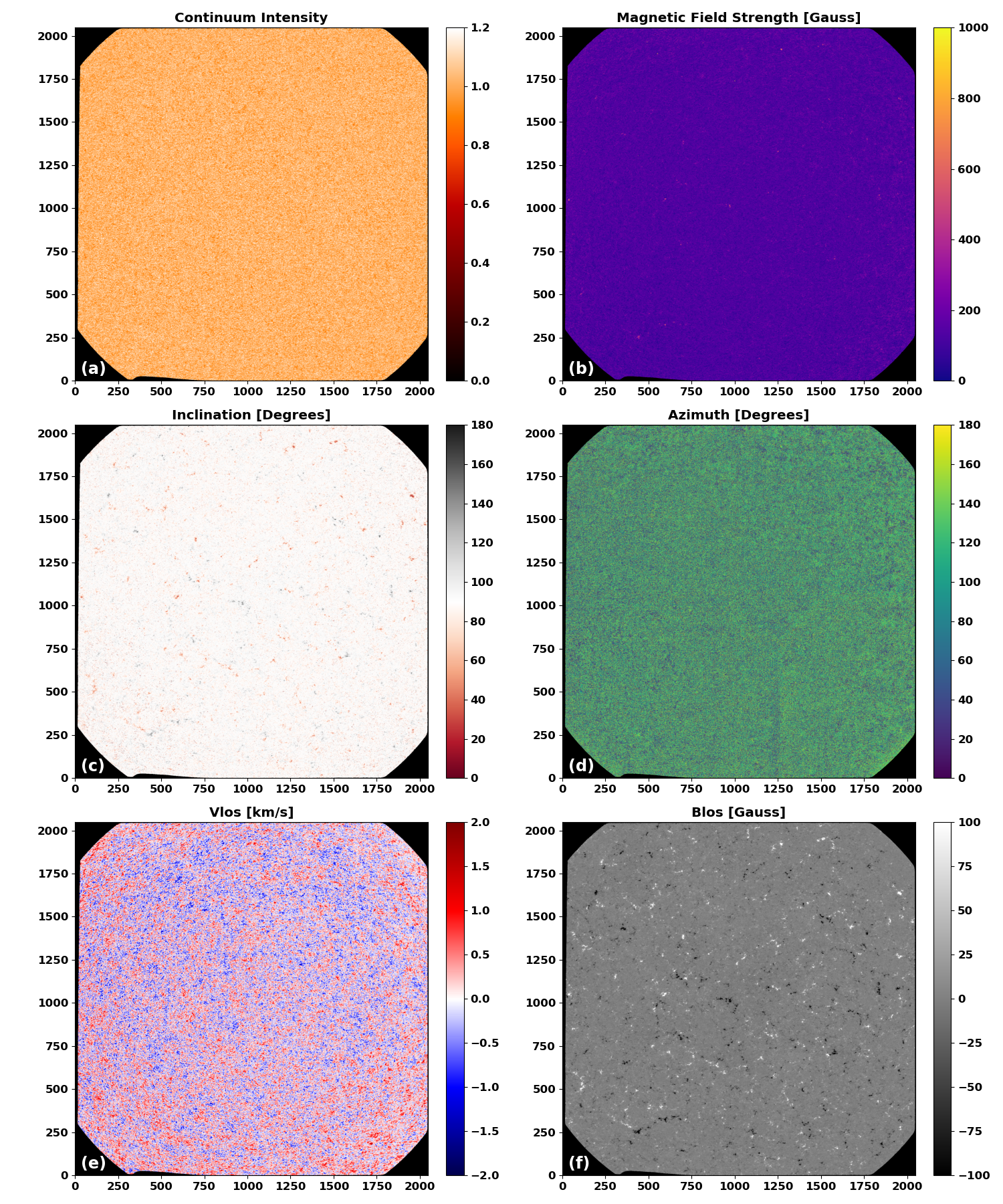}
\caption[rte] 
{ \label{fig:rte}Derived physical quantities from data of 23 February 2021 17:00 UTC: a) continuum intensity, b) magnetic field strength, c) inclination relative to line-of-sight, d) azimuth relative to direction of rotation, e) Dopplergram, f) line-of-sight magnetogram. The region outside the field stop is set to black for clarity.}
\end{center}
\end{figure}

The Stokes filtergrams in Fig. \ref{fig:iquv} display high uniformity and low linear polarimetric signal as expected for a quiet Sun. This demonstrates the high effectiveness of the flat-field correction and additional cross-talk removal. The photospheric magnetic field network appears clearly in the Stokes $V/I_{c}$ image. In Stokes $U/I_{c}$ the remnants of a polarimetric ghost edge is present in the lower right corner. This ghost is most likely due to a reflection off the HREW (see Sec. \ref{subsec:flat_corr}). The flat-field correction removes the ghost to a large extent but the edge remains. From analysis of histograms of the four quadrants, the difference of the lower right corner distribution from the others is below the $10^{-3}\cdot I_{c}$ noise level. Figure \ref{fig:rte} displays the derived quantities from these Stokes filtergrams.

As expected by the uniformity of the filtergrams, the physical quantities in Fig. \ref{fig:rte} display equal uniformity and low magnetic field strengths due to the quiet Sun being void of active regions. The edge of a polarimetric ghost is visible in the lower right corner of the azimuth due to the absence of signal. The continuum intensity map exhibits the granular structure of the photosphere. The inclination is centred on $90$ degrees, and due to the very low linear polarisation signal, the azimuth contains mainly noise. 

\begin{figure} [H]
\begin{center}
\includegraphics[width = \textwidth]{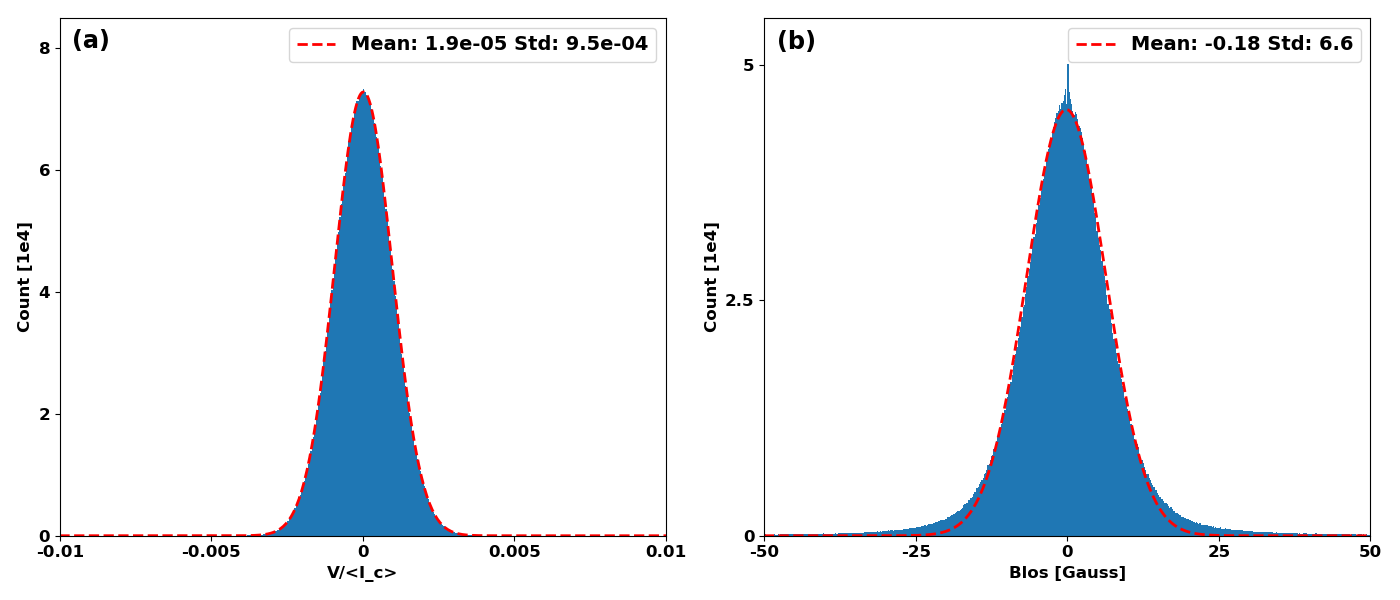}
\caption[blos noise] 
{ \label{fig:noise}Histograms from data on 23 February 2021 1700 UTC: a) Stokes $V/I_{c}$ at the continuum wavelength position, with a Gaussian fit (dashed red curve), b) the line-of-sight magnetogram, with a Gaussian fit.}
\end{center}
\end{figure}

The Gaussian fit to the Stokes $V$ histogram in Fig. \ref{fig:noise} a) indicates that a polarimetric accuracy of $<10^{-3}\cdot I_{c}$ is achieved, illustrating the high performance of the HRT instrument. It is also important to note that due to the tight telemetry budget, raw images from SO/PHI are compressed before download. The compression procedure, in this case to $6$ bits/pixel (down from $32$), increases the noise of the filtergrams: for example, data from the commissioning phase, which was downloaded without compression, had a Stokes noise level of $8.5\times10^{-4}$. Furthermore, using the same method as Liu et al. (2012)\cite{liu_comparison_2012}, the line-of-sight magnetogram has an estimated noise level of $6.6$ G, very similar to the noise level of the $720$ second magnetogram images from HMI, but with almost eight times the cadence: $96$ seconds.

\subsection{November 2021}
\begin{figure} [H]
\begin{center}
\includegraphics[width = \textwidth]{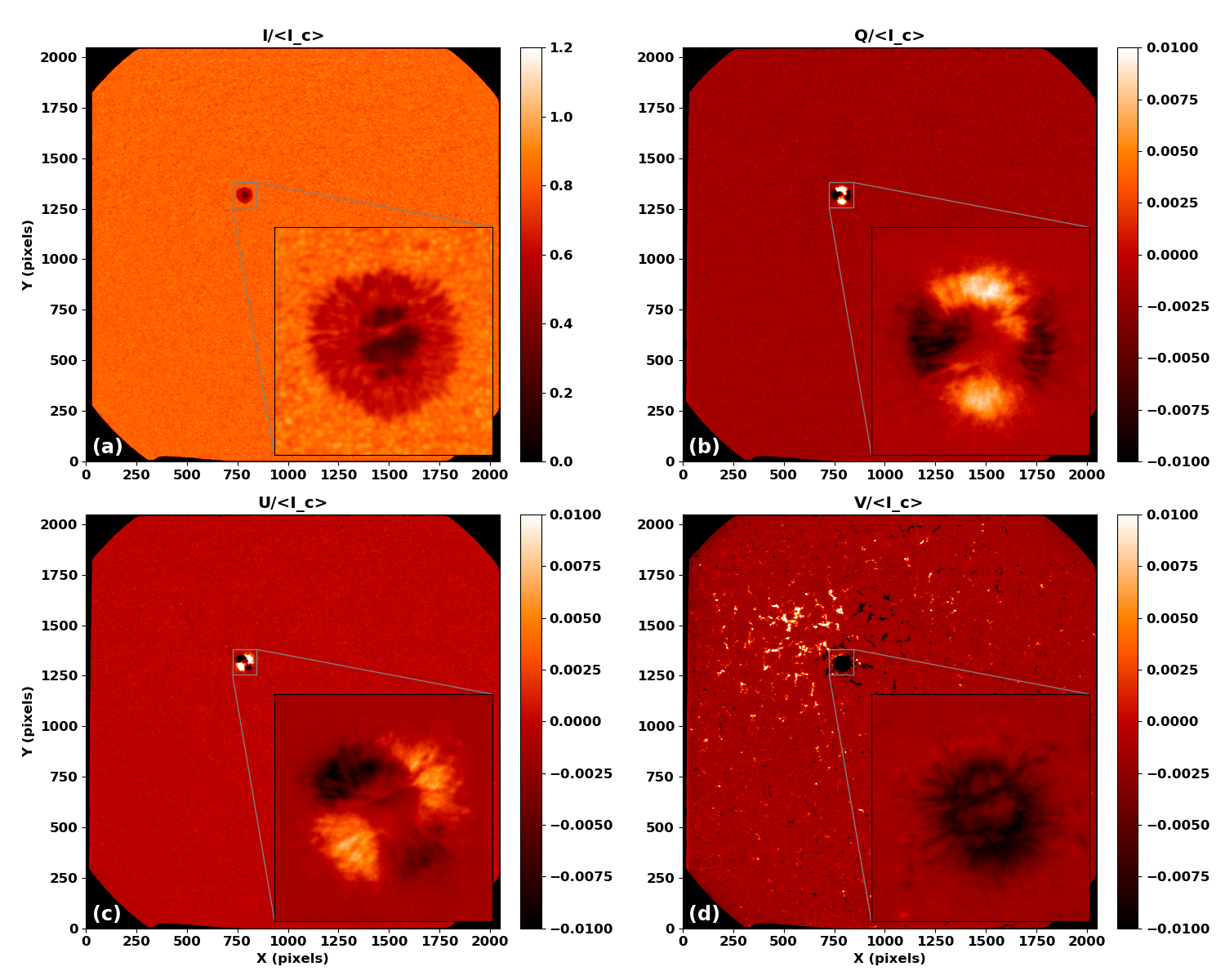}
\caption[stokesnov] 
{ \label{fig:iquv_nov}Stokes filtergram in the red line wing on 5 November 2021 20:21 UTC: a) Stokes $I/I_{c}$ b)  Stokes $Q/I_{c}$ c)  Stokes $U/I_{c}$ d)  Stokes $V/I_{c}$.}
\end{center}
\end{figure}

We present a reduced dataset of a sunspot captured by HRT, taken during the inferior conjunction in November 2021. The spacecraft was flying close to Earth, with a distance to the Sun of $0.858$ AU, a PMP temperature set to $40\;\degree$C and was pointing to disc centre.  At this distance the (two-pixel) spatial resolution is $624$ km, and almost half the solar disk is within the FOV. As shown in Fig. \ref{fig:iquv_nov} there are clear signals in Stokes $Q/I_{c}$ and $U/I_{c}$ that capture the linear polarisation from the sunspot, which highlights the instrument's sensitivity. The $45$ degree offset in the signal pattern between Stokes $Q/I_{c}$ and $U/I_{c}$ is also finely highlighted.

\begin{figure} [H]
\begin{center}
\includegraphics[width = \textwidth]{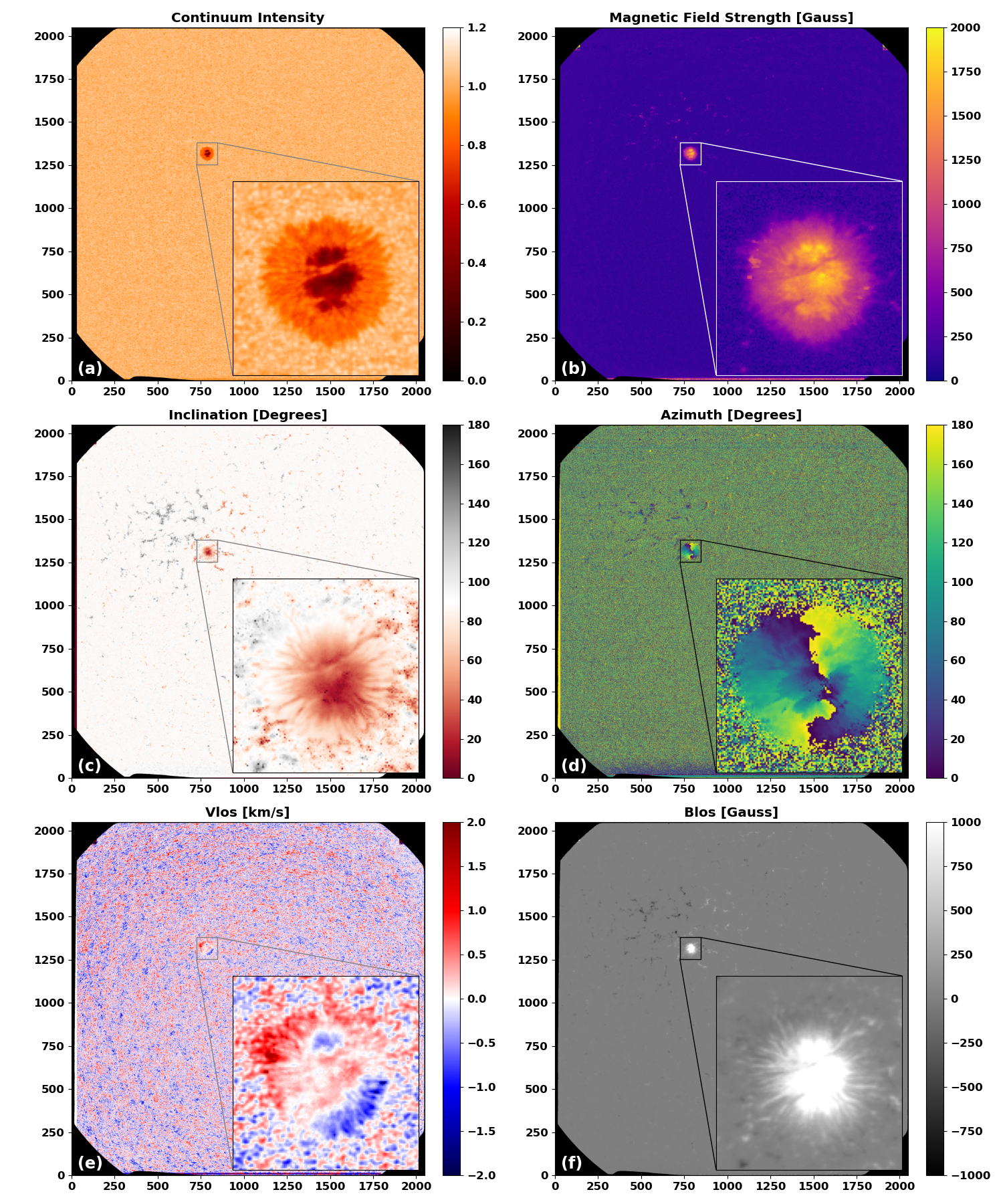}
\caption[rtesunspot] 
{ \label{fig:rte_nov}Derived physical quantities from data of 5 November 2021 20:21 UTC : a) continuum intensity, b) magnetic field strength, c) inclination relative to line-of-sight, d) azimuth relative to direction of rotation, e) Dopplergram, f) line-of-sight magnetogram. The region outside the field stop is set to black for clarity.}
\end{center}
\end{figure}

Figure \ref{fig:rte_nov} displays the physical quantities computed from the Stokes filtergrams plotted in Fig. \ref{fig:iquv_nov}. Selecting the umbra region with a continuum upper threshold of $0.6$, the mean magnetic field strength in the umbra is $1420$ G. This is somewhat low for an umbra and may reflect straylight, or that the large Zeeman splitting within the umbra is not caught that well by the placement of the wavelength points in PHI. The azimuth is of particular interest with a strong signal. The line-of-sight velocity displays the expected redshift on the limb side of the spot, with the corresponding blueshift towards disc centre (Evershed flow)\cite{Evershed1909RadialSun-spots}.

\subsection{Magnetogram Noise}
\begin{figure} [H]
\begin{center}
\includegraphics[width = \textwidth]{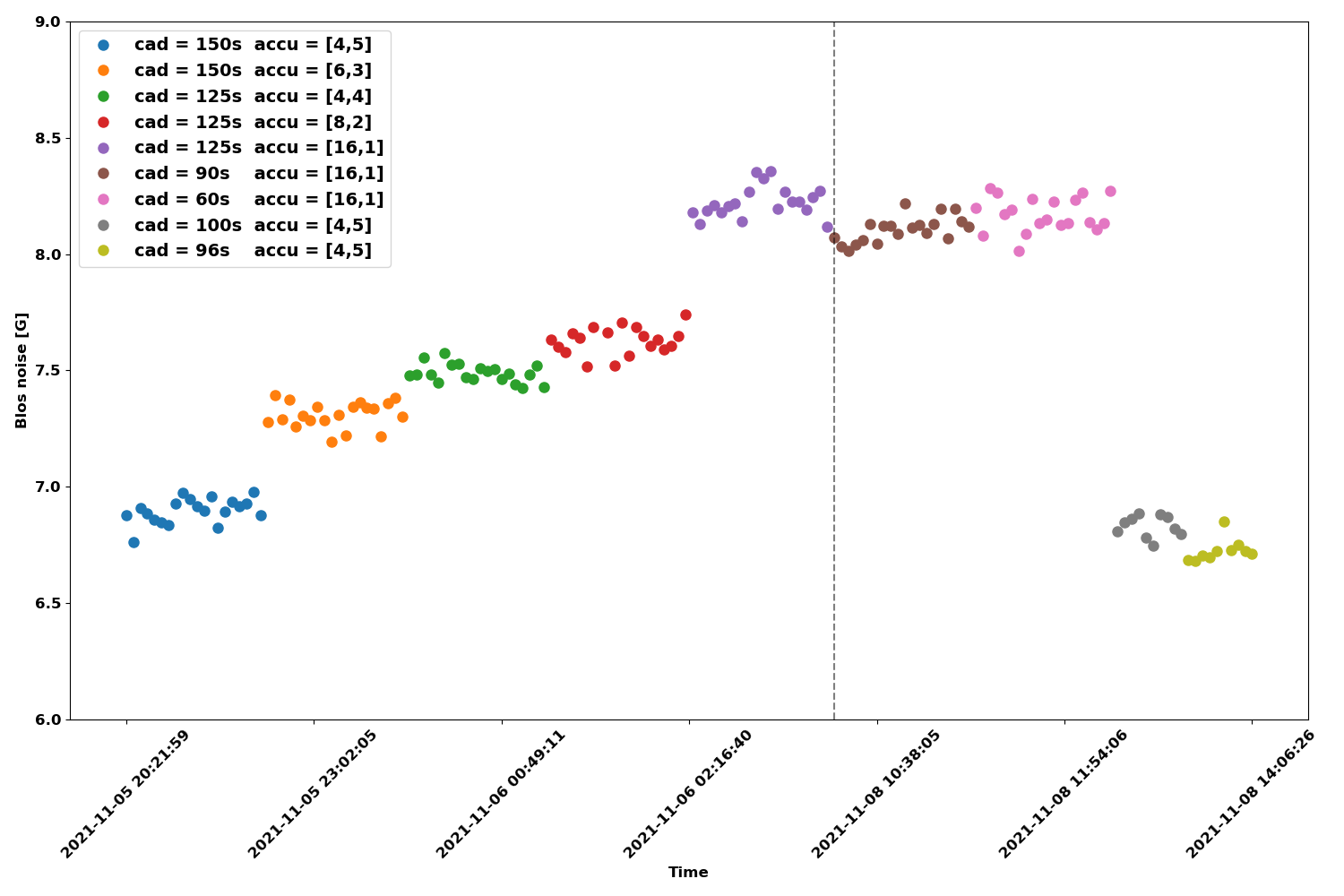}
\caption[cadnoise] 
{ \label{fig:cad_noise} Noise from the central region of the magnetogram plotted against time. The different combinations of `cad' (cadence) and `accu' (PMP modulation accumulation scheme, see Sec. \ref{subsec:hrt_inst}) are denoted by their marker colours. The vertical dashed line signifies the change in time from November 6th to November 8th.}
\end{center}
\end{figure}

Several datasets with different modulation schemes were acquired during a campaign in early November 2021 to test the cadence of the schemes. The noise from the line-of-sight magnetograms of these datasets at different cadences is presented in Fig. \ref{fig:cad_noise}. The total number of frames per image is found from the multiplication of the two numbers in the accumulation scheme. A clear trend is visible: as the accumulation scheme changes from $[4,5]$ to $[16,1]$, less frames are being accumulated for each image, and therefore as expected the magnetogram noise increases from $6.8$ G to $8.3$ G. The last grouping, was the fastest the $[4,5]$ scheme could be executed by the instrument, with a cadence of $96$ seconds. It is also clear that the higher the cycles of the modulation states (the second value in the accumulation scheme), the lower the magnetogram noise. It must also be noted that like the data from February 2021, the compression acts as the main driver of the noise.

\section{Conclusion}\label{sec:conc}
An on-ground pipeline has been developed to reduce raw data from the HRT instrument to produce high quality data with a polarimetric accuracy $10^{-3}\cdot I_{c}$ and infer physical parameters from the polarised light. The $96$ second cadence line-of-sight magnetograms are shown to have an excellent low level of noise, only $6.6$ G, similar to the noise level of the HMI $720$ second magnetograms. This was achieved by calibrating the flat-fields to remove unwanted artefacts from the acquisition process by use of unsharp masking. As a a result of the analysis presented here, the unsharp masking procedure will be implemented onboard the spacecraft, such that the in-flight data will also produce data of the highest quality. %
The absence of the ISS has also been taken care by three more steps. Despite the increase in data quality, as shown in Fig. \ref{fig:mod_align}, noise levels remain slightly higher than in the standard configuration because of the spacecraft jitter. %
This pipeline will be embedded into a software tool which will automatically process all the SO/PHI science data that will arrive on ground and store them on the appropriate databases.
\acknowledgments 
 
This work was carried out in the framework of the International Max Planck Research School (IMPRS) for Solar System Science at the Max Planck Institute for Solar System Research (MPS). Solar Orbiter is a space mission of international collaboration between ESA and NASA, operated by ESA.  We are grateful to the ESA SOC and MOC teams for their support. The German contribution to SO/PHI is funded by the BMWi through DLR and by MPG central funds. The Spanish contribution is funded by FEDER/AEI/MCIU (RTI2018-096886-C5), a “Center of Excellence Severo Ochoa” award to IAA-CSIC (SEV-2017-0709), and a Ramón y Cajal fellowship awarded to DOS. The French contribution is funded by CNES. 
\bibliography{new_med.bib} 

\begin{thebibliography}{10}

\bibitem{Muller2020TheOverview}
M{\"{u}}ller, D., St.~Cyr, O.~C., Zouganelis, I., Gilbert, H.~R., Marsden, R.,
  Nieves-Chinchilla, T., Antonucci, E., Auch{\`{e}}re, F., Berghmans, D.,
  Horbury, T.~S., Howard, R.~A., Krucker, S., Maksimovic, M., Owen, C.~J.,
  Rochus, P., Rodriguez-Pacheco, J., Romoli, M., Solanki, S.~K., Bruno, R.,
  Carlsson, M., Fludra, A., Harra, L., Hassler, D.~M., Livi, S., Louarn, P.,
  Peter, H., Sch{\"{u}}hle, U., Teriaca, L., Del Toro~Iniesta, J.~C.,
  Wimmer-Schweingruber, R.~F., Marsch, E., Velli, M., De~Groof, A., Walsh, A.,
  and Williams, D., ``{The Solar Orbiter mission - Science overview},'' {\em
  Astronomy {\&} Astrophysics}~{\bf 642},  A1 (10 2020).

\bibitem{solanki_polarimetric_2020}
Solanki, S.~K., Del Toro~Iniesta, J.~C., Woch, J., Gandorfer, A., Hirzberger,
  J., Alvarez-Herrero, A., Appourchaux, T., Mart{\'{i}}nez~Pillet, V.,
  P{\'{e}}rez-Grande, I., Sanchis~Kilders, E., Schmidt, W., G{\'{o}}mez~Cama,
  J.~M., Michalik, H., Deutsch, W., Fernandez-Rico, G., Grauf, B., Gizon, L.,
  Heerlein, K., Kolleck, M., Lagg, A., Meller, R., M{\"{u}}ller, R.,
  Sch{\"{u}}hle, U., Staub, J., Albert, K., Alvarez~Copano, M., Beckmann, U.,
  Bischoff, J., Busse, D., Enge, R., Frahm, S., Germerott, D., Guerrero, L.,
  L{\"{o}}ptien, B., Meierdierks, T., Oberdorfer, D., Papagiannaki, I.,
  Ramanath, S., Schou, J., Werner, S., Yang, D., Zerr, A., Bergmann, M.,
  Bochmann, J., Heinrichs, J., Meyer, S., Monecke, M., M{\"{u}}ller, M.~F.,
  Sperling, M., {\'{A}}lvarez~Garc{\'{i}}a, D., Aparicio, B.,
  Balaguer~Jim{\'{e}}nez, M., Bellot~Rubio, L.~R., Cobos~Carracosa, J.~P.,
  Girela, F., Hern{\'{a}}ndez~Exp{\'{o}}sito, D., Herranz, M., Labrousse, P.,
  L{\'{o}}pez~Jim{\'{e}}nez, A., Orozco~Su{\'{a}}rez, D., Ramos, J.~L.,
  Barandiar{\'{a}}n, J., Bastide, L., Campuzano, C., Cebollero, M.,
  D{\'{a}}vila, B., Fern{\'{a}}ndez-Medina, A., Garc{\'{i}}a~Parejo, P.,
  Garranzo-Garc{\'{i}}a, D., Laguna, H., Mart{\'{i}}n, J.~A., Navarro, R.,
  N{\'{u}}{\~{n}}ez~Peral, A., Royo, M., S{\'{a}}nchez, A., Silva-L{\'{o}}pez,
  M., Vera, I., Villanueva, J., Fourmond, J.~J., De~Galarreta, C.~R., Bouzit,
  M., Hervier, V., Le~Clec'h, J.~C., Szwec, N., Chaigneau, M., Buttice, V.,
  Dominguez-Tagle, C., Philippon, A., Boumier, P., Le~Cocguen, R., Baranjuk,
  G., Bell, A., Berkefeld, T., Baumgartner, J., Heidecke, F., Maue, T., Nakai,
  E., Scheiffelen, T., Sigwarth, M., Soltau, D., Volkmer, R.,
  Blanco~Rodr{\'{i}}guez, J., Domingo, V., Ferreres~Sabater, A., Gasent~Blesa,
  J.~L., Rodr{\'{i}}guez~Mart{\'{i}}nez, P., Osorno~Caudel, D., Bosch, J.,
  Casas, A., Carmona, M., Herms, A., Roma, D., Alonso, G., G{\'{o}}mez-Sanjuan,
  A., Piqueras, J., Torralbo, I., Fiethe, B., Guan, Y., Lange, T., Michel, H.,
  Bonet, J.~A., Fahmy, S., M{\"{u}}ller, D., and Zouganelis, I., ``{The
  Polarimetric and Helioseismic Imager on Solar Orbiter},'' {\em Astronomy and
  Astrophysics}~{\bf 642},  A11 (10 2020).

\bibitem{gandorfer_high_2018}
Gandorfer, A., Grauf, B., Staub, J., Bischoff, J., Woch, J., Hirzberger, J.,
  Solanki, S.~K., {\'{A}}lvarez-Herrero, A., Parejo, P.~G., and Schmidt, W.,
  ``{The High Resolution Telescope (HRT) of the Polarimetric and Helioseismic
  Imager (PHI) onboard Solar Orbiter},'' in [{\em Space Telescopes and
  Instrumentation 2018: Optical, Infrared, and Millimeter
  Wave}{\nolinebreak\hspace{0.1em}]},   {\bf 10698},  106984N, International
  Society for Optics and Photonics (2018).

\bibitem{Lange2017On-boardInstrument}
Lange, T., Fiethe, B., Michel, H., Michalik, H., Albert, K., and Hirzberger,
  J., ``{On-board processing using reconfigurable hardware on the solar orbiter
  PHI instrument},'' {\em 2017 NASA/ESA Conference on Adaptive Hardware and
  Systems, AHS 2017} ,  186--191 (9 2017).

\bibitem{Albert2020}
Albert, K., Hirzberger, J., Kolleck, M., Jorge, N.~A., Busse, D.,
  Rodr{\'{i}}guez, J.~B., Carrascosa, J. P.~C., Fiethe, B., Gandorfer, A.,
  Germerott, D., Guan, Y., Guerrero, L., Gutierrez-Marques, P., Exp{\'{o}}sito,
  D.~H., Lange, T., Michalik, H., Su{\'{a}}rez, D.~O., Schou, J., Solanki,
  S.~K., Iniesta, J. C. d.~T., and Woch, J., ``{Autonomous on-board data
  processing and instrument calibration software for the Polarimetric and
  Helioseismic Imager on-board the Solar Orbiter mission},'' {\em
  https://doi.org/10.1117/1.JATIS.6.4.048004}~{\bf 6},  048004 (12 2020).

\bibitem{Zouganelis2020}
Zouganelis, I., De~Groof, A., Walsh, A.~P., Williams, D.~R., M{\"{u}}ller, D.,
  St~Cyr, O.~C., Auch{\`{e}}re, F., Berghmans, D., Fludra, A., Horbury, T.~S.,
  Howard, R.~A., Krucker, S., Maksimovic, M., Owen, C.~J.,
  Rodr{\'{i}}guez-Pacheco, J., Romoli, M., Solanki, S.~K., Watson, C., Sanchez,
  L., Lefort, J., Osuna, P., Gilbert, H.~R., Nieves-Chinchilla, T., Abbo, L.,
  Alexandrova, O., Anastasiadis, A., Andretta, V., Antonucci, E., Appourchaux,
  T., Aran, A., Arge, C.~N., Aulanier, G., Baker, D., Bale, S.~D., Battaglia,
  M., Bellot~Rubio, L., Bemporad, A., Berthomier, M., Bocchialini, K., Bonnin,
  X., Brun, A.~S., Bruno, R., Buchlin, E., B{\"{u}}chner, J., Bucik, R.,
  Carcaboso, F., Carr, R., Carrasco-Bl{\'{a}}zquez, I., Cecconi, B.,
  Cernuda~Cangas, I., Chen, C.~H., Chitta, L.~P., Chust, T., Dalmasse, K.,
  D'Amicis, R., Da~Deppo, V., De~Marco, R., Dolei, S., Dolla, L., Dudok De~Wit,
  T., Van Driel-Gesztelyi, L., Eastwood, J.~P., Espinosa~Lara, F., Etesi, L.,
  Fedorov, A., F{\'{e}}lix-Redondo, F., Fineschi, S., Fleck, B., Fontaine, D.,
  Fox, N.~J., Gandorfer, A., G{\'{e}}not, V., Georgoulis, M.~K., Gissot, S.,
  Giunta, A., Gizon, L., G{\'{o}}mez-Herrero, R., Gontikakis, C., Graham, G.,
  Green, L., Grundy, T., Haberreiter, M., Harra, L.~K., Hassler, D.~M.,
  Hirzberger, J., Ho, G.~C., Hurford, G., Innes, D., Issautier, K., James,
  A.~W., Janitzek, N., Janvier, M., Jeffrey, N., Jenkins, J., Khotyaintsev, Y.,
  Klein, K.~L., Kontar, E.~P., Kontogiannis, I., Krafft, C., Krasnoselskikh,
  V., Kretzschmar, M., Labrosse, N., Lagg, A., Landini, F., Lavraud, B., Leon,
  I., Lepri, S.~T., Lewis, G.~R., Liewer, P., Linker, J., Livi, S., Long,
  D.~M., Louarn, P., Malandraki, O., Maloney, S., Martinez-Pillet, V.,
  Martinovic, M., Masson, A., Matthews, S., Matteini, L., Meyer-Vernet, N.,
  Moraitis, K., Morton, R.~J., Musset, S., Nicolaou, G., Nindos, A., O'Brien,
  H., Orozco~Suarez, D., Owens, M., Pancrazzi, M., Papaioannou, A., Parenti,
  S., Pariat, E., Patsourakos, S., Perrone, D., Peter, H., Pinto, R.~F.,
  Plainaki, C., Plettemeier, D., Plunkett, S.~P., Raines, J.~M., Raouafi, N.,
  Reid, H., Retino, A., Rezeau, L., Rochus, P., Rodriguez, L.,
  Rodriguez-Garcia, L., Roth, M., Rouillard, A.~P., Sahraoui, F., Sasso, C.,
  Schou, J., Sch{\"{u}}hle, U., Sorriso-Valvo, L., Soucek, J., Spadaro, D.,
  Stangalini, M., Stansby, D., Steller, M., Strugarek, A.,
  {{\v{S}}tver{\'{a}}k}, Susino, R., Telloni, D., Terasa, C., Teriaca, L.,
  Toledo-Redondo, S., Del Toro~Iniesta, J.~C., Tsiropoula, G., Tsounis, A.,
  Tziotziou, K., Valentini, F., Vaivads, A., Vecchio, A., Velli, M., Verbeeck,
  C., Verdini, A., Verscharen, D., Vilmer, N., Vourlidas, A., Wicks, R.,
  Wimmer-Schweingruber, R.~F., Wiegelmann, T., Young, P.~R., and Zhukov, A.~N.,
  ``{The Solar Orbiter Science Activity Plan - Translating solar and
  heliospheric physics questions into action},'' {\em Astronomy {\&}
  Astrophysics}~{\bf 642},  A3 (10 2020).

\bibitem{Alvarez-Herrero2015a}
Alvarez-Herrero, A., Garc{\'{i}}a~Parejo, P., Laguna, H., Villanueva, J.,
  Barandiar{\'{a}}n, J., Bastide, L., Reina, M., S{\'{a}}nchez, A., Gonzalo,
  A., Navarro, R., Vera, I., and Royo, M., ``{Polarization modulators based on
  liquid crystal variable retarders for the Solar Orbiter mission},'' in [{\em
  Polarization Science and Remote Sensing VII}{\nolinebreak\hspace{0.1em}]},
  {\bf 9613}(September 2015),  96130I (2015).

\bibitem{Volkmer2012}
Volkmer, R., Bosch, J., Feger, B., Gomez, J.~M., Heidecke, F., Schmidt, W.,
  Scheiffelen, T., Sigwarth, M., and Soltau, D., ``{Image stabilisation system
  of the photospheric and helioseismic imager},'' in [{\em Space Telescopes and
  Instrumentation 2012: Optical, Infrared, and Millimeter
  Wave}{\nolinebreak\hspace{0.1em}]},   {\bf 8442},  84424P, SPIE (9 2012).

\bibitem{Carmona2014}
Carmona, M., G{\'{o}}mez, J.~M., Roma, D., Casas, A., L{\'{o}}pez, M., Bosch,
  J., Herms, A., Sabater, J., Volkmer, R., Heidecke, F., Maue, T., Nakai, E.,
  and Schmidt, W., ``{System model of an image stabilization system},'' in
  [{\em Modeling, Systems Engineering, and Project Management for Astronomy
  VI}{\nolinebreak\hspace{0.1em}]},   {\bf 9150},  91501U, SPIE (8 2014).

\bibitem{Dominguez-Tagle2014OpticalPHI}
Dominguez-Tagle, C., Appourchaux, T., Ruiz~de Galarreta, C., Fourmond, J.-J.,
  Philippon, A., Le~Clec'h, J.-C., Bouzit, M., Bommier, V., Le~Cocguen, R.,
  Crussaire, D., and Malherbe, J.-M., ``{Optical characterization of the
  breadboard narrowband prefilters for Solar Orbiter PHI},'' {\em Space
  Telescopes and Instrumentation 2014: Optical, Infrared, and Millimeter
  Wave}~{\bf 9143},  91435G (8 2014).

\bibitem{kuhn_gain_1991}
Kuhn, J.~R., Lin, H., and Loranz, D., ``{Gain calibrating nonuniform
  image-array data using only the image data},'' {\em Publications of the
  Astronomical Society of the Pacific}~{\bf 103}(668),  1097 (1991).

\bibitem{DelToroIniesta2003IntroductionSpectropolarimetry}
Del Toro~Iniesta, J.~C.,  [{\em {Introduction to
  Spectropolarimetry}}{\nolinebreak\hspace{0.1em}]}, Cambridge University Press
  (3 2003).

\bibitem{SanchezAlmeida1992ObservationSunspots}
Sanchez~Almeida, J., Lites, B.~W., Sanchez~Almeida, J., and Lites, B.~W.,
  ``{Observation and Interpretation of the Asymmetric Stokes Q, U, and V Line
  Profiles in Sunspots},'' {\em ApJ}~{\bf 398},  359 (10 1992).

\bibitem{schlichenmaier_spectropolarimetry_2002}
Schlichenmaier, R. and Collados, M., ``{Spectropolarimetry in a sunspot
  penumbra. Spatial dependence of Stokes asymmetries in Fe I 1564.8 nm},'' {\em
  Astronomy and Astrophysics}~{\bf 381},  668--682 (1 2002).

\bibitem{Guizar-Sicairos2008}
Guizar-Sicairos, M., Thurman, S.~T., and Fienup, J.~R., ``{Efficient subpixel
  image registration algorithms},'' {\em Optics Letters}~{\bf 33},  156 (1
  2008).

\bibitem{hoeksema_helioseismic_2014}
Hoeksema, J.~T., Liu, Y., Hayashi, K., Sun, X., Schou, J., Couvidat, S.,
  Norton, A., Bobra, M., Centeno, R., Leka, K.~D., Barnes, G., and Turmon, M.,
  ``{The Helioseismic and Magnetic Imager (HMI) Vector Magnetic Field Pipeline:
  Overview and Performance},'' {\em Solar Physics}~{\bf 289},  3483--3530 (9
  2014).

\bibitem{Borrero2011}
Borrero, J.~M., Tomczyk, S., Kubo, M., Socas-Navarro, H., Schou, J., Couvidat,
  S., and Bogart, R., ``{VFISV: Very Fast Inversion of the Stokes Vector for
  the Helioseismic and Magnetic Imager},'' {\em Solar Physics}~{\bf 273},
  267--293 (2 2011).

\bibitem{LandiDeglInnocenti2005}
Landi~Degl'Innocenti, E. and Landolfi, M.,  [{\em {Polarization in spectral
  lines}}{\nolinebreak\hspace{0.1em}]}, Kluwer Academic Publishers (2005).

\bibitem{orozco_suarez_usefulness_2007}
Su{\'{a}}rez, D.~O. and Del Toro~Iniesta, J.~C., ``{The usefulness of analytic
  response functions},'' {\em Astronomy and Astrophysics}~{\bf 462},
  1137--1145 (2 2007).

\bibitem{CobosCarrascosa2016}
Cobos~Carrascosa, J.~P., Aparicio~del Moral, B., Ramos~Mas, J.~L., Balaguer,
  M., L{\'{o}}pez~Jim{\'{e}}nez, A.~C., and del Toro~Iniesta, J.~C., ``{The RTE
  inversion on FPGA aboard the solar orbiter PHI instrument},'' {\em Software
  and Cyberinfrastructure for Astronomy IV}~{\bf 9913},  991342 (7 2016).

\bibitem{Semel1967}
Semel, M., ``{Contribution {\`{a}} l'{\'{e}}tude des champs magn{\'{e}}tiques
  dans les r{\'{e}}gions actives solaires},'' {\em Annales
  d'Astrophysique}~{\bf 30},  513--551 (1967).

\bibitem{Rees1978}
Rees, D. E.;~Semel, M.~D., ``{Line formation in an unresolved magnetic
  element-A test of the centre of gravity method},'' {\em Astronomy and
  Astrophysics}~{\bf 74}(4),  1--5 (1979).

\bibitem{liu_comparison_2012}
Liu, Y., Hoeksema, J.~T., Scherrer, P.~H., Schou, J., Couvidat, S., Bush,
  R.~I., Duvall, T.~L., Hayashi, K., Sun, X., and Zhao, X., ``{Comparison of
  Line-of-Sight Magnetograms Taken by the Solar Dynamics
  Observatory/Helioseismic and Magnetic Imager and Solar and Heliospheric
  Observatory/Michelson Doppler Imager},'' {\em Solar Physics}~{\bf 279},
  295--316 (7 2012).

\bibitem{Evershed1909RadialSun-spots}
Evershed, J., ``{Radial movement in sun-spots},'' {\em The Observatory}~{\bf
  32},  291--292 (1909).

\end{thebibliography}
\bibliographystyle{spiebib} 

\end{document}